\documentclass[12pt]{article}

\usepackage{epsfig}
\usepackage{dcolumn}
\begin{document}
\baselineskip=.165in

\newcommand{\be}{\begin{equation}}
\newcommand{\ee}{\end{equation}}
\newcommand{\bea}{\begin{eqnarray}}
\newcommand{\eea}{\end{eqnarray}}
\newcommand{\da}{\dagger}
\newcommand{\dg}[1]{\mbox{${#1}^{\dagger}$}}
\newcommand{\hlf}{\mbox{$1\over2$}}
\newcommand{\lfrac}[2]{\mbox{${#1}\over{#2}$}}
\newcommand{\scsz}[1]{\mbox{\scriptsize ${#1}$}}
\newcommand{\tsz}[1]{\mbox{\tiny ${#1}$}}
\newcommand{\ch}       {\mbox{\bf (*** CHECK! ***)~}}
\renewcommand{\descriptionlabel}[1]%
		{\hspace{\labelsep}\textsf{#1}}

\begin{flushright} 
gr-qc/0308017 \\
LA-UR-03-5540 \\
\end{flushright} 

\begin{flushleft}

\large{\bf Finding the Origin of the Pioneer Anomaly}

\vspace{0.5in}

\normalsize
\bigskip 

Michael Martin Nieto${^a}$ and  Slava G. Turyshev$^b$\\ 

\normalsize
\vskip 15pt

${^a}$Theoretical Division (MS-B285), Los Alamos National Laboratory,\\
University of California,  Los Alamos, New Mexico 87545, U.S.A. \\
Email: mmn@lanl.gov \\ 
\vspace{0.25in}
$^{b}$Jet Propulsion Laboratory, California Institute of  Technology,\\
Pasadena, CA 91109, U.S.A. \\ 
Email: turyshev@jpl.nasa.gov


\vspace{0.5in}
\bigskip

\end{flushleft}

\baselineskip=.165in

\begin{abstract}

Analysis of radio-metric tracking data from the Pioneer 10/11
spacecraft at distances between 20 - 70 astronomical units (AU) from
the Sun has consistently indicated the presence of an anomalous, 
small, constant Doppler frequency drift. The drift can be
interpreted as being due to a constant acceleration of $a_P= (8.74 \pm
1.33) \times 10^{-8}$  cm/s$^2$ directed {\it towards} the Sun. 
Although it is suspected that there is a systematic origin to the
effect, none has been found.  As a result, the nature of this anomaly
has become of growing interest.  Here we present a concept for a deep-space
experiment that will reveal the origin of the discovered anomaly and
also will characterize its properties to an accuracy of 
at least two orders of magnitude below the anomaly's size.   
The proposed mission will not only provide a
significant accuracy improvement in the search for small anomalous
accelerations, it will also determine if the anomaly is due to some
internal systematic or has an external origin.  A number of critical
requirements and design considerations for the mission are outlined
and addressed. If only already existing technologies were used,
the mission could be flown as early as 2010.  
\\

\end{abstract}

\begin{center}
\today
\end{center}

\newpage


\section{The Pioneer Missions and the Anomaly}
\label{intro}
	
The Pioneer 10/11 missions, launched on 2 March 1972 (Pioneer
10) and 4 Dec. 1973 (Pioneer 11), were the first to explore
the outer solar system \cite{old1}. 
After Jupiter and (for Pioneer
11) Saturn encounters, the two spacecraft followed escape
hyperbolic orbits near the plane of the ecliptic to opposite
sides of the solar system. Pioneer 10 eventually became the
first man-made object to leave the solar system.   
	
By 1980, when Pioneer 10 passed a distance of  $\sim$ 20 AU from the
Sun, the acceleration contribution from solar-radiation pressure on
the craft (directed away from the Sun) had decreased to less than $4
\times 10^{-8}$ cm/s$^2$.  At that time the navigational data had
already indicated the presence of an anomaly in the Doppler data; but
at first the anomaly was only considered to be an interesting navigational
curiosity and was not seriously analyzed.   
	
This changed in 1994 when, since the anomaly had not disappeared, 
an inquiry was initiated into its possible origin \cite{bled}.  
The consequence was a long-term collaboration to study and understand
the Pioneer data in hand.  Useful data were recorded almost up to the
end of the official Pioneer 10 mission in 2001, with the last signal
from the spacecraft being received on 22 January 2003 \cite{pioend}.    

The initial results of the study were reported in 1998 \cite{old2} and
a detailed analysis appeared in 2002 \cite{old3}.  For this final
analysis the existing Pioneer 10/11 Doppler data from 1987.0
to 1998.5 was used \cite{old3}. Realizing the potential significance of the
discovery, all {\it known} sources of a
possible systematic origin for the detected anomaly 
were specifically addressed.  However,  even
after all {\it known} systematics were accounted for, the 
conclusion remained that there was an anomalous acceleration signal of 
$a_P=(8.74 \pm 1.33) \times 10^{-8}$  cm/s$^2$ in the direction towards the
Sun.   This anomaly was a constant with respect to both time and
distance, for distances between about 20 to 70 AU from the Sun.

We emphasize {\it known} because one might naturally expect that there 
is a systematic origin to the effect, perhaps generated by the
spacecraft themselves from excessive heat or propulsion gas
leaks.  But neither we nor others with spacecraft or navigational
expertise have been able to find a convincing explanation for such a
mechanism \cite{old2}-\cite{old4}. This inability to explain
the anomalous acceleration of the Pioneer spacecraft with conventional
physics has contributed to the growing discussion about its origin. 

Attempts to verify the anomaly using other spacecraft proved 
disappointing. This is because the Voyager, Galileo, Ulysses, and
Cassini spacecraft navigation data
all have their own individual difficulties for use in an independent
test of the anomaly. (See Sections \ref{stabile} and \ref{hyperbolic}.) 
In addition, many of the deep space missions that are currently being
planned either will not provide the needed navigational accuracy and
trajectory stability of under $10^{-8}$ cm/c$^2$ (i.e., Pluto Express) 
or else they will have significant on-board systematics (see
Sec. \ref{on-board}) that 
mask the anomaly (i.e., JIMO -- Jupiter Icy Moons Orbiter). 

To enable a clean test of the anomaly there is also a requirement 
to have an escape hyperbolic trajectory. 
(See Sec.~\ref{sec:lessons} for more details.)
This makes a number of other missions (i.e., LISA -- the Laser
Interferometric Space Antenna, STEP -- Space Test of Equivalence
Principle,  LISA Pathfinder, etc.) less able  
to directly test the anomalous acceleration.  Although these missions
all have excellent scientific goals and technologies, nevertheless,
because of their orbits they will be in a less advantageous position
to conduct a precise test of the detected anomaly. 

Thus, the origin of this anomaly remains unclear.  No 
unambiguous ``smoking gun'' on-board systematic has been found
\cite{old4}.  This can be seen by the number of theoretical ideas for
new physics that have been proposed to explain the anomaly.\footnote{
For a review and summary up to the start of 2002, see
Section XI of \cite{old3}.}  
By way of illustration, we give two examples.  
A drag force by enough ``dark" mirror-matter  
could cause the acceleration \cite{foot}. The acceleration 
due to drag from {\it any kind of} interplanetary medium is 
$
a_{\tt d}(r)= -{\mathcal{K}}_{\tt d}~\rho(r)~v_{s}^2(r)~A/m, 
$
where 
$\mathcal{K}_{\tt d}$ is the effective
reflection/absorption/transmission coefficient of the 
spacecraft surface being hit, $\rho(r)$ is the density of the
interplanetary medium,  
$v_{s}(r)$ is the effective
relative velocity of the craft to the medium, $A$ is the cross-section
of the craft, and $m$ is its mass.  A constant density (which would be
hard to understand) of  
$4 \times 10^{-19}$ g/cm$^3$ would therefore explain the Pioneer
anomaly \cite{foot}.\footnote{  
We observe that this argument also holds for
normal matter, although there is no evidence for that high a constant
density \cite{old3}.}
Another idea is Modified Newtonian Dynamics
(MOND)  \cite{old7}, which describes a situation where $F \sim 1/r$ at
large distances and which has 
an acceleration parameter similar to $a_P$. It has been noted that for
a hyperbolic orbit like the Pioneers' there will be an anomalous
acceleration similar to ours  \cite{old7}. 

With this background, we assert that one can no longer ignore the
signal and it is time to experimentally settle the issue
with a new deep-space mission that will 
test for and decide  the origin of the anomaly \cite{old5,bertolami}.  
Here we propose such a new mission, one that would enable an
independent and unambiguous test of the Pioneer anomaly and also   
improve the accuracy of its determination. 

When proposing any space mission one needs to address two 
important issues: the scientific justification for the
mission objectives;  and  the mission configuration and design
requirements, including the overall construction, launch, and ground
operations cost. 

Our arguments above show that there is 
a strong scientific justification to fly a mission to discover the
origin of the Pioneer anomaly.  Therefore, 
in Section \ref{concept} we proceed to the mission
issues.  We review the lessons learned from the Pioneer 10/11
spacecraft, explain the spin-stabilization, on-board power, the 
``fore/aft''\footnote{DSN radio tracking convention has that
when a Pioneer antenna points toward the Earth, this 
defines the ``aft'', ``backward'', or ``rear''  
direction on the spacecraft.  The equipment
compartment placed on the other side of of the antenna defines the
``fore'', ``forward'', or ``front'' direction on the spacecraft.} 
symmetric bus and antenna designs, the hyperbolic escape
orbit, and the launch concept. 
In Section \ref{navigation} we discuss the navigation plan, 
the data that will be obtained, and the precision with which
one can characterize it.  
Section \ref{sec:systematics} describes the systematics and 
the error budget for the mission's fundamental goal --
the small acceleration signal. 
We close with a summary. 

Our test is
designed to unambiguously determine whether or not the  anomaly is due
to some unknown physics or else to an on-board systematic.  
As pointed out in the literature \cite{old5,bertolami}, 
either way the result would be of major significance. 
If the anomaly is a manifestation of new or
unexpected physics, the result would be of truly fundamental
importance. However, even if the anomaly turns out to be a
manifestation of an unknown systematic mechanism,
understanding it will affect the way small forces will be handled in
future precision space navigation.


\section{Mission Concept}
\label{concept}

\subsection{Applying Lessons from the Pioneers}
\label{sec:lessons}

\begin{figure}[t!]
 \begin{center}
\noindent    
\psfig{figure=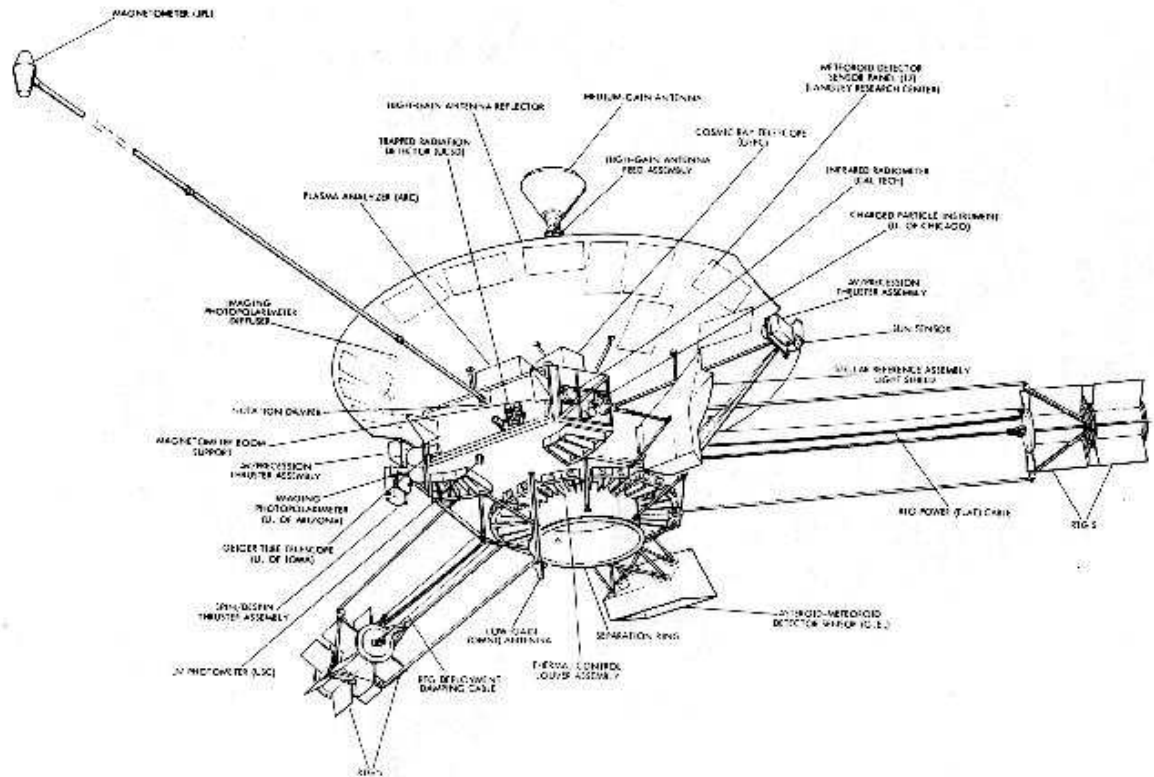,width=155mm}
\end{center}
  \caption{A drawing of the Pioneer spacecraft.  
 \label{fig:trusters}}
\end{figure} 


Our experience, studying both the Pioneer anomaly and also 
the Pioneer craft itself, 
has given us much insight into how to design a test of the anomaly.
Previously we begun to develop a concept for
a mission for such an investigation \cite{old5}, and  
we have continued to identify a number of critical design
requirements.  In this section we will
discuss these requirements, their significance, and our approach to
addressing them. 

To explain what configuration and design will be best 
for the proposed mission, it first is helpful 
to summarize  what made the Pioneer craft work so well. (See Figure 
\ref{fig:trusters}.)  Among the most important features the Pioneers
had were  \cite{old3}:  
(i) simple attitude control realized with a spin-stabilized architecture; 
(ii) on-board nuclear power sources;
(iii) a well-understood thermal control system; 
(iv) deep-space, hyperbolic, escape-orbit trajectories; and 
(v) extensive navigational coverage with high accuracy Doppler tracking. 

Our goal is to design a mission using only existing technology
that would ensure a spacecraft environment with systematics 
reduced to the order of ${0.1 \times 10^{-8}}$ cm/s$^2$ or less. 
That is, we want to have a spacecraft with the geometry and the 
associated physical properties that will allow a definition of the major
elements of mission operations such that unknown sources of
non-gravitational accelerations affecting the
spacecraft's motion will be reduced to unprecedented levels. 
The spacecraft and mission design outlined in the
following subsections directly respond to this
stated goal and will allow minimization of the contributions of various
known systematic errors. 


\subsection{Spacecraft Stabilization}
\label{stabile} 

For deep-space navigational purposes the Pioneer spacecraft were much
easier to navigate than any other spacecraft, including 
the Voyagers,\footnote{
The Voyagers are three-axis stabilized.  The resulting oft-used gas
jets yield a navigation error of $\sim 10^{-6}$ cm/s$^2$, which is an
order of magnitude larger than the Pioneer anomaly \cite{old3}.} 
Galileo,\footnote{
Galileo was only spin-stabilized during Earth-Jupiter cruise.  Although
this data set was useful to verify the Deep Space Network hardware, it
came from so close in to the Sun that it was too highly correlated with
the solar radiation pressure to yield a conclusive result \cite{old3}.}
Ulysses,\footnote{Ulysses had to have an  excessive number of maneuvers due
to a failed nutation damper.  Although the analysis was indicative,
individual errors were as much as an order of magnitude larger than the
effect \cite{old3}.} 
and Cassini.\footnote{The Cassini craft used reaction wheels 
during Jupiter-Saturn cruise, obviating precise navigation 
modeling \cite{jda}. There also was a large effect from the 
RTGs being mounted on the end of the craft.}
This was achieved by utilizing a simple
spin-stabilized spacecraft architecture -- the two Pioneers were {\it
always} simple spinners.  

When in deep space (say at 20 AU or greater), spin-stabilized spacecraft
like the Pioneers require only a single maneuver every few 
months or so to correct for the drift of the antennae pointing
direction due to the effect of the craft's proper motion. 
Thus, the Pioneers had no continuous- or often-utilized-jetting 
of attitude control gas.  This would have made the navigational 
accuracy too poor, as happened with the 3-axis-stabilized Voyagers.
This is one of the main reasons the Pioneers were so well tracked.  

Further, 
modern 3-axis stabilization relies heavily on the use of precise
fuel gauges (to measure fuel usage during maneuvers for input
into navigational models), high quality thrusters (for precise
attitude control purposes), reaction wheels (to keep preferred
spacecraft pointing for a limited time), and often high resolution
accelerometers (to track on-board generated non-gravitational
disturbances). 
Although there exist fuel gauges with the desirable
precision, thrusters have low  repeatability and reaction wheel
de-saturation introduces high acceleration noise.\footnote{ 
This was the case for the state-of-the-art propulsion 
assemblies used for attitude control of the Cassini mission. 
That mission has only $\sim 40\%$ truster repeatability and 
significant reaction  wheel noise \cite{jda}.} 
Finally, existing pico-g level accelerometers  also have 
low reliability.  This all makes 3-axis stabilization a very costly
and undesirable choice for our deep space mission.  

Existing spin-stabilized attitude control technology (including
in-space propulsion modules, fuel gauges, and even thrusters -- because
they are seldom used), enables orbit determination precise to better
than\footnote{Indeed, even the Pioneers had a data precision almost
as good as this. 
The size of the effect could be determined well.  It was
determining the contributions making up the anomaly (the systematics
vs. a ``true'' signal) that was the problem.}
${\sim 0.003 \times 10^{-8}}$ cm/s$^2$ (see Section
\ref{sec:methodology}), more than two orders of magnitude smaller than
the level of the error in the Pioneer anomaly.
Therefore, when considering the new mission architecture,
we prefer a spin-stabilized spacecraft as opposed to one that is
3-axis stabilized.  


\subsection{On-board Energy Source}
\label{energy} 

Because this will be a deep-space mission, the distance and time
involved rule out solar cells or batteries.  An autonomous 
nuclear source is the only viable option for power. 
There are currently two choices for such an energy source, 
a nuclear-electric propulsion module (similar to the one that is being
considered for the JIMO mission) and Radioisotope Thermoelectric
Generators (RTGs).   

The use of a nuclear-electric propulsion module would definitely solve
the power problem by providing virtually unlimited power for 
the scientific instrument.  But it would also make precise navigation
and any related navigational science investigation a difficult task.
(See Section \ref{heatsymmetry}.) 
In any event, nuclear-electric propulsion modules are still being
developed and have not yet been flown.  Therefore, although they do
have a very strong potential for deep space exploration in the future,
for now they are not a viable solution. They may become the basic
propulsion elements and sources for on-board power for missions that
will fly 20 years from now. 

On the other hand, RTGs are the present conventional source of
energy for any deep space mission. 
By having the RTGs on extended booms, that are deployed
after launch, one obtains the  rotational stability of the craft
discussed above and also gets a reduction in the heat systematics. 
(See Section \ref{heatsymmetry}.) 
This is why, for this new mission, we choose the use of RTGs with a
mounting approach similar to that of the Pioneers.\footnote{
In passing we note that if any  
new missions like this fly, then the Plutonium itself will likely 
come from Russia with the safety testing and analysis, fuel
purification, and heat source fabrication done in the United States.
This could inspire international and intra-agency cooperation
on the program, since independently there is revived interest in RTGs
and in nuclear-electric propulsion  \cite{old6}.}


\subsection{Heat-Symmetric Spacecraft Design}  
\label{heatsymmetry} 

For a nuclear powered spacecraft, perhaps the major navigation systematic 
in deep space is thermal emission generated by the spacecraft's power system.  
This is because, with either space-craft centered RTGs or else
a space-craft centered nuclear reactor, 
there are many to hundreds of kW of heat power generated.  This 
also produces at least hundreds of W of electrical power in the bus.  
The heat dissipation can produce a non-isotropic force on the craft
which can dominate a force of size the Pioneer anomaly, especially if
the craft is light.   For example,  only $\sim 63$ W of directed power 
could have explained the anomaly the 241 kg Pioneer craft with half
its fuel depleted. 
Therefore, if the RTGs had been placed ``forward''
they obviously would have yielded a huge systematic

We will eliminate the heat systematic by making
the heat dissipation fore/aft symmetric.  In a stroke of serendipitous 
luck,\footnote{Because they were 
the first deep space craft, the Pioneer engineers were worried
about the effects of nuclear radiation on the main bus electronics.
Placing the RTGs far away at the end of booms was the solution.}
the Pioneer RTGs, with  $\sim$ 2,500 W of heat, were placed at the end
of booms.  This meant they had little thermal effect on the
craft. Further, the rotation of the 
Pioneer craft and their RTG fin structure design meant the
radiation was extremely symmetrical fore-aft, with very little heat
radiated in the direction towards the craft. The same concept will be
used for this mission, with perhaps shielding of the RTGs to further
prevent anisotropic heat reflection.   

The electrical power in the equipment and instrument compartments must
also be radiated out so as not to cause an undetected systematic.  For
the Pioneers the central compartment was surrounded by insulation.  
There were louvers forward to be open and 
let out excess heat early in the mission and to be closed and retain
heat later on when the electrical power was less. The electrical power
degrades faster than the radioactive decay because 
{\it any} degradation of the thermoelectric
components means the electric power degrades from this {\it on top of}
the degradation of the input heat due to the 87.74
half-life of the $^{238}$Pu.\footnote{
For the Pioneers, the time from launch by when the Pioneer 10 electrical
power had been reduced to 50\% was about 20 years.\cite{old4}}
For this mission, the louvers will 
be on the side of the compartment so they radiate in an axially
symmetric manner as the spacecraft rotates.  

\begin{figure}[t] 
    \noindent
    \begin{center}  
        \begin{minipage}[t]{.46\linewidth} 
            \epsfig{file=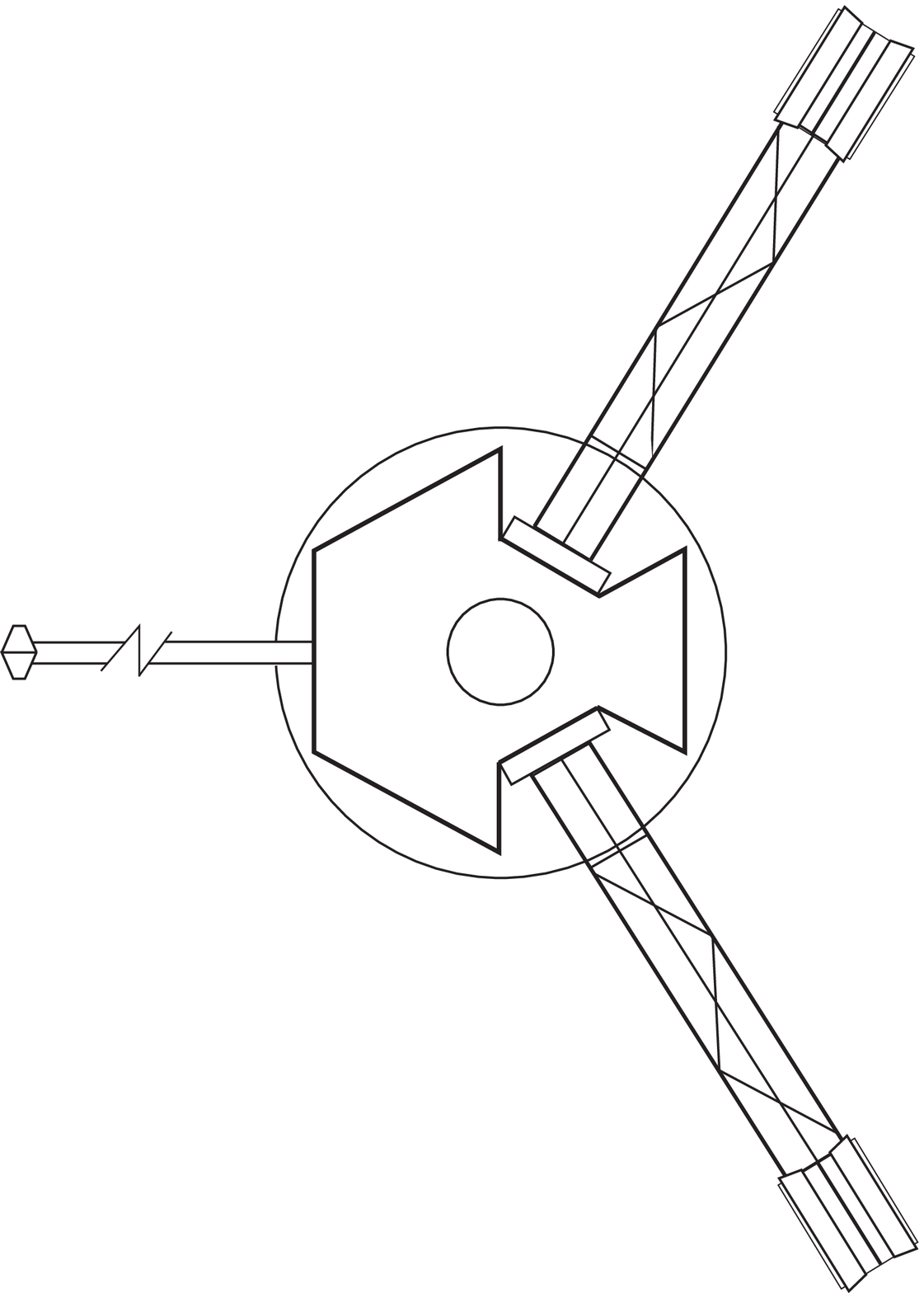,width=70mm}
            \noindent
        \end{minipage}
        \hskip 15pt
        \begin{minipage}[t]{.46\linewidth}
            \epsfig{file=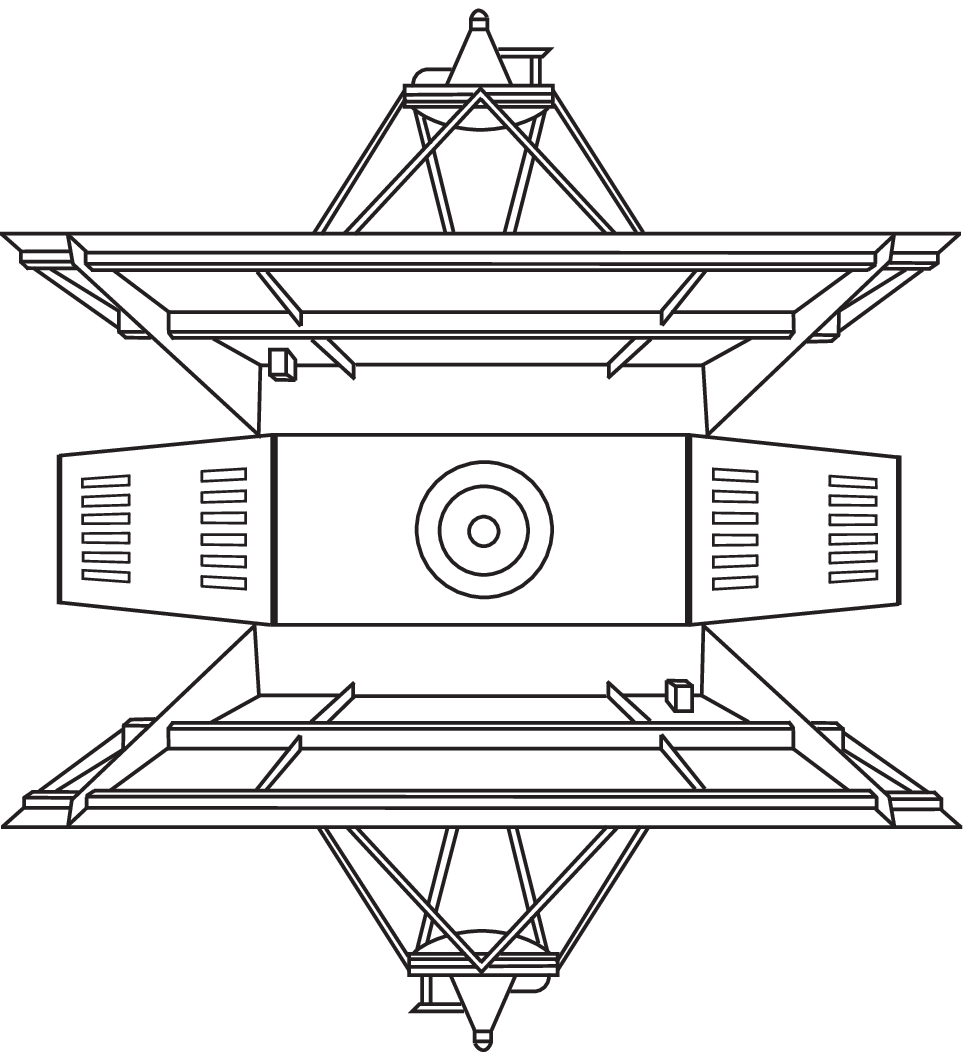,width=70mm}
            \noindent
        \end{minipage}
\caption{The top (left) and side (right) views (different scales)
of our ``yo-yo" craft concept. The scale of the circular antenna is on
the order of 2 to 2.5 m.   The RTGs are deployed on the left.  There
also is an indication of a third long boom where an instrument 
package to detect interstellar matter could be placed. 
Depending on the final mission objectives this instrument
package could be replaced by a third RTG. The side view
shows the louvers radiating to the side and the antennas, taken and
modeled from the Cassini Cassegrain antenna \cite{casantenna}.}
            \label{yoyo}  
    \end{center}
\end{figure}

In Figure \ref{yoyo} we show a concept design.  Although
unconventional, a unique feature of our concept is the dual,
identical, fore/aft antenna system. A spacecraft design such as this
has never been proposed before. However, this symmetric deign
allows us further to minimize the heat systematic.  Any heat
reaching the back of the two antennas. despite insulation placed
in, around, and in the support of the bus, will be reflected
symmetrically fore/aft.

Our preliminary analysis (see Sec. \ref{on-board}) 
suggests that with the existing technologies one can balance the
fore/aft geometry of the spacecraft to minimize the possible
differential heat rejection systematic to the level  
${\leq 0.03 \times 10^{-8}}$ cm/s$^2$.   Thus, this  
fore/aft symmetric design greatly reduces the size of any 
possible heat systematic.\footnote{
For the Pioneers, contributions to the detected anomaly of
order  $10^{-8}$ cm/s$^2$ 
came individually from the RTGs and power dissipation \cite{old3}.
(See Section \ref{sec:systematics}.)} 

A final factor in the spacecraft heat transfer mechanism 
that we want to mention is the
optical properties of the spacecraft surfaces. This is a 
challenging issue to discuss quantitatively. 
The difficulty lies in the precise folding of the
reflective insulation blankets and in the precision painting of all the
external surfaces. 
Of course, for our fore/aft design we will make the processes very symmetric.  
But it is still hard to predict the exact behavior of all the
surfaces on the spacecraft after launch, especially after long
exposure to the space environment (i.e., solar radiation, dust,
planetary fly-byes, etc.). 
However, this did not seem to affect the Pioneer results
\cite{old3,old4}, and this mission's use of
rotating the antennas (described in 
the next subsection) will obviate any residual effect. 

\subsection{More Symmetry: Identical fore/aft antennas}  
\label{antennas}

One great advantage of the dual, identical, fore/aft antenna system
(shown in Figure  \ref{yoyo}) is the ability to significantly reduce
the effect of the recoil force from the radio-communication beam.\footnote{
It was 8~W for the Pioneers, which contributed a bias of  $\sim 1.1
\times 10^{-8}$ cm/s$^2$ to the detected acceleration \cite{old3}.} If
the signal is continuously beamed in 
both directions, the beam radiation reaction will cancel to at least  
${\sim 0.01 \times 10^{-8}}$ cm/s$^2$ (depending on the quality of the
design and the components used). This is because the beam force in
either direction will be $\sim 1 \times 10^{-8}$ cm/s$^2$, and a
quality control of 1\% on the antennas would yield this limit.  
Thus, there would be no need to account
for this systematic.  

Given that the antenna is on the scale of 2--2.5 m (the Pioneers had
``9 ft'' = 2.74 m antennas) and that there is a similar layout to the
Pioneers, {\it except}
for the added second antenna, one would expect this craft to be around
300 kg or less.  (At launch the Pioneers {\it with} hydrozene fuel
weighed 259 kg.)

But most importantly, after one has determined a precise signal with
one orientation (perhaps after a year or two of data taking) then,
aided by Sun and star sensors, the craft can be rotated by 180
degrees so the forward antenna will then be backwards and vice
versa.\footnote{A very similar rotation, the ``Earth Acquisition Maneuver,'' 
was actually performed on Pioneer 10 soon after launch.  For a craft
like the Pioneers such a maneuver can be done in about two hours and
take about 0.5 kg of fuel.}   Then, if the anomaly is 
due to an external effect the measurement will remain the same after
rotation whereas the force would be in the opposite direction after
rotation if it were due to an on-board systematic.    
A different (non-zero)
result in the two orientations would also unambiguously demonstrate that 
their was both (a) an externally caused anomaly (one-half the sum of the two
measurements) combined with (b) an internal systematic (one-half the
difference of the two results).  

Therefore, this unique
``yo-yo" design will yield an {\it unambiguous test} of whether the anomaly 
is due to an internal systematic or to some unknown external origin. 
It is a major element of the mission concept.  


\subsection{Hyperbolic Escape Orbits}
\label{hyperbolic}

The Pioneer anomaly was found on craft following hyperbolic, un-bound, 
escape trajectories.  
Contrariwise, solar system data tells us that the
anomaly is not seen in the trajectories of large bodies that are bound
in low eccentricity orbits.  Objects with larger eccentricities, such
as long period comets, do show evidence of anomalous behavior, but
the significant mass loss masks any signal at the order of the Pioneer
anomaly.   For the various experimental reasons 
mentioned in Section \ref{stabile}, the 
indicative data from Ulysses and Galileo in cruise was too 
noisy to be used to draw any conclusion.
There also exist anomalies seen in hyperbolic planetary
flybys.\footnote{Anomalous energy increases have been observed in 
Earth flybys; for example in Galileo's first flyby in 1990 
\cite{old-flyby, flyby} and the NEAR flyby in 1998  \cite{flyby}.  
There may also have been a small anomaly in
the 1999 Cassini flyby \cite{new-flyby}.}  This all emphasizes how 
the transition from bound to escape orbits has
never been well characterized \cite{new-flyby}. 

The anomaly was precisely measured between 20 and 70 AU out from the
Sun.  Although, it might have been present closer in, this has only  
been imprecisely studied \cite{old4}. For this reason and also 
to reduce the effect of external systematics the experiment should reach
distances greater than 15 AU from the Sun.  Obviously, one wants the
time needed to reach this region to be short; say, not much more than
6 years.   
To yield a direct test for any velocity-dependence in the signal, 
one also wants the craft to have a significantly different velocity than
the Pioneers.  

All this means that when the craft reaches deep space it should be in
a high-velocity, hyperbolic, escape orbit. 


 \subsection{The Launch Vehicle}
\label{sec:launch}

For a successful mission, the above spacecraft requirements have to
be integrated with a launch concept (and also with a total scientific
package if there are other experiments).

The launch vehicle is a major consideration for any deep-space mission.
To test the Pioneer anomaly cleanly, one wants to reach a distance
greater than 20 AU to be able to clearly distinguish any effect from 
solar radiation pressure and other near-solar systematics. The craft
is projected to be of small mass (say, 300 kg or less).\footnote{ 
The Pioneers weighed 259 kg at launch. Therefore, with a second
antenna, 300 kg is a bound which should be improved by using modern
materials.} Even so, because the desired distance is large (from less
than 20 AU to as much as greater than 70 AU) a large solar system
escape velocity is desired (say, more than 10 AU/yr).  In contrast,  
the Pioneers are cruising at a velocity of about 2 AU/year and the
Voyagers at about 3 AU/year. One needs something faster than that.   

The obvious first idea is a very energetic rocket.\footnote{The Russian
Proton rocket has a very successful record.  Using it is an intriguing
possibility. Indeed, this again might be a useful option for
international collaboration to hold down the cost to NASA or ESA.
Further, when the Atlas V, Delta IV,  and Ariane V are fully developed,
they will provide other potential vehicles.  Since launch will be no 
earlier than 2010, this question can be carefully considered.}
If a rocket is the source of the large velocity,
then a test of Pioneer anomaly might be performed as the radio-science 
objective of, say, a mission to the outer solar system.  We would 
integrate as many as possible of our design criteria into 
the constraints of the main mission.  

Better yet would be to have our experiment 
be independent and jettisoned after final
powered acceleration so it could fly on alone.  This would eliminate any
cross-talk systematics from the main mission.  A related 
possibility would be having our mission piggy-back on a
large craft having nuclear power as the basis for acceleration, 
such as the ion engine of the Prometheus program \cite{prometheus}.  
Here our craft would remain attached to the mother craft until a suitable
velocity was reached, say  10 AU/yr.  Then our craft
would separate to allow our described program of testing for a small force.

It is also interesting, as a speculative alternative, to consider a symbiotic
relationship to a solar sail mission. Both NASA \cite{sail1}-\cite{sail4}
and ESA\footnote{Through the German Space Agency, ESA is considering a
sail for deep-space travel as a development of the earlier Odissee
concept \cite{germanyesa}.}  \cite{germanyesa,dachwald}
are developing solar sails. The NASA InterStellar Probe mission
concept \cite{sail1}-\cite{sail4}  would reach the boundaries of
interstellar space, the termination shock, the heliopause ($>$ 150
AU), and the bow shock, all expected to be well past 100 AU.  The sail
would be jettisoned beyond Jupiter.  This mission foresees sending a
package (of about our size and configuration) at a speed of 14
AU/year.   This velocity would be ideal for us and such a mission
could also be combined with using a solar sail to detect matter in the 
ecliptic by measuring the drag force produced \cite{saildrag}.


\section{Navigation Plan}
\label{navigation}

Even if all systematics were known, no good radio-science data set is
possible without good navigation. This implies the use of modern
navigational techniques, such as both Doppler 
and range radio-tracking methods (and perhaps others discussed below
\cite{marsrover}).  

Doppler measures the velocity
of the craft. It only indirectly yields a distance to the
craft when one integrates the measured Doppler velocity from known initial 
conditions.  Range itself is a time-of-flight measurement. This is
done by phase modulating the signal and timing the return signal 
which was transponded at the craft.  As such, it gives the distance to
the spacecraft directly and is a complementary check in the orbit
determination.  Therefore, range will yield an independent and hence
very precise test of the Pioneer anomaly.\footnote{The Pioneers had
only Doppler communication capabilities, making impossible any
independent verification of the anomaly with a range signal
\cite{old3}.} 

An added precision will be available with the occasional use of   
Very Long Baseline Interferometry (VLBI) that enables the 
differenced Doppler ($\Delta$DOR) technique. $\Delta$DOR is similar to
ranging, but it also takes in a third signal from a naturally
occurring radio source in space, such as a quasar.  This additional
source helps scientists and engineers gain a more accurate location of
the spacecraft.\footnote{
This ``ranging'' is not really ranging, but differenced ranging. What
is measured is the difference in the distances to the source from 
two DSN complexes on Earth (for example, Goldstone and Madrid or
Goldstone and Canberra). From that an angle in the sky can be
determined relative to the stations. The angle for the quasar is
subtracted from the angle of the spacecraft, giving the angular
separation of the quasar and the spacecraft. That angle is accurate to
about five to ten nanoradians.  This means that if a spacecraft is
near Mars, say 200 million kilometers away, the
position of the spacecraft can be determined to within one
kilometer. Recently this technique was 
successfully used in navigating the two Mars missions that 
carried the rovers Spirit 
and Opportunity \cite{marsrover}.}  

The mission capabilities might also include the use of
multi-frequency communication.  This is because multi-frequency
communication is useful for  correcting dispersive media effects.  In 
particular, it allows the precise calibration of solar and
interplanetary plasma systematics. This is why in future missions
it would be useful to utilize more than one among the S- ($\sim$2.4
GHz), X- ($\sim$7.2 GHz), or Ka- ($\sim$32.3 GHz) frequency bands.   
In our discussion below we will concentrate on the use of X-band,
because it is presently the standard technology for radio-science, 
and Ka-band, because it is well en-route to being a standard.  

A difficult problem in deep-space navigation is precise 3-dimensional
orbit determination.  The ``line-of-sight'' component of a velocity is 
much more easily determined by Orbit Determination Program (ODP) codes
than are the motions in the orthogonal directions. But having both
Doppler and range can mitigate this. With the precise, low-systematic
data and the analysis of it that we are calling for, a much better
than usual determination can be made of the orthogonal dynamics, even
to the point of obtaining good three-dimensional acceleration solutions. 
Additionally, $\Delta$DOR observations will further reduce the
uncertainty in the plane-of-the-sky components of the spacecraft
proper motion. These navigational capabilities will enable an anomaly
test with a sensitivity below ${0.003 \times10^{-8}}$ cm/s$^2$ for
distances in the range 20 to 90 AU (see Sec. \ref{sec:methodology}). 

\begin{figure}[t!]
 \begin{center}
\noindent   
\psfig{figure=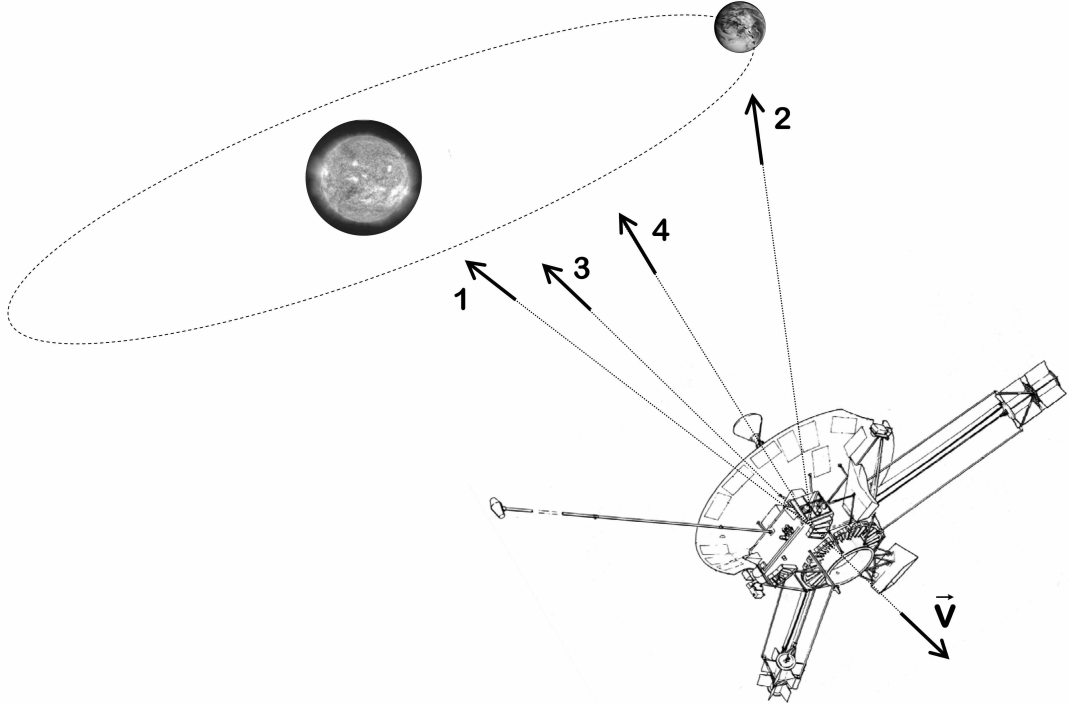,width=145mm}
\end{center}
\vskip -10pt 
  \caption{Four possible directions for the anomalous acceleration
acting on the Pioneer spacecraft:  (1) towards the Sun, (2)
towards the Earth, (3) along the direction of motion of the craft,
or (4) along the spin axis.
 \label{fig:direction}}
\end{figure} 

\begin{figure}[t!]
 \begin{center}
\noindent   
\psfig{figure=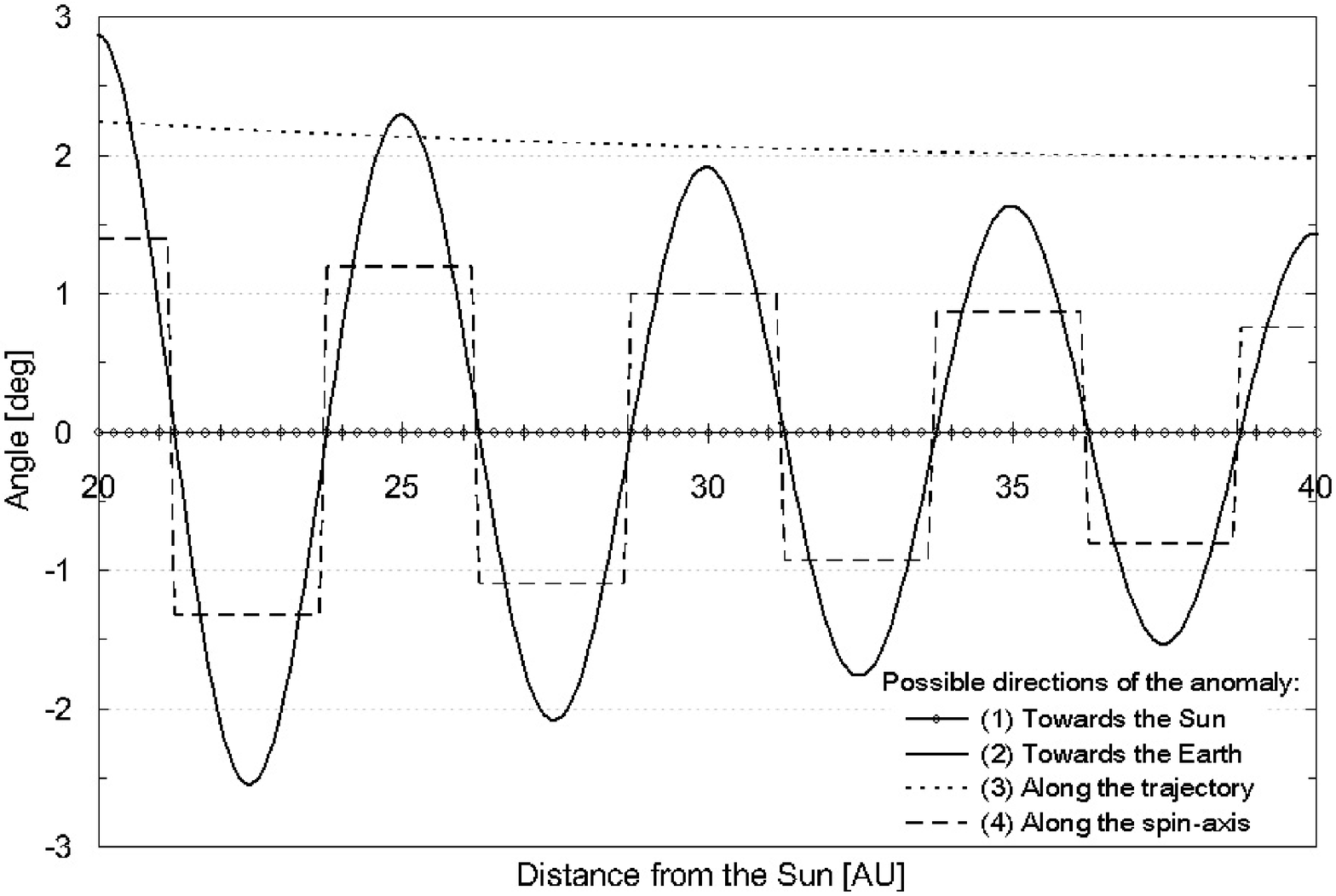,width=145mm}
\end{center}
\vskip -15pt 
  \caption{The signatures for four possible directions of the
anomalous acceleration acting on the proposed spacecraft. The
signatures are distinctively different and are easily detectable with
the proposed mission. (See the text.) 
 \label{fig:angles}}
\end{figure} 


With the radiation pattern of the Pioneer antennae and the lack of 
precise 3-D navigation, the determination 
of the exact direction of the anomaly was a difficult task \cite{old3}. 
For standard antennae, and without good 3-D navigation, 
in deep space the directions (1) towards the Sun, (2)
towards the Earth, (3) along the direction of motion of the craft,
or (4) along the spin axis, are all observationally synonymous. 
These directions (see Figure \ref{fig:direction}) would tend to
indicate an origin that is 
(1) new dynamical physics originating from the Sun, 
(2) a time signal anomaly, 
(3) a drag or inertial effect, or 
(4) an on-board systematic. 

At 20 AU these directions are of order 3 degrees apart (the
maximum angle subtended by the Sun and the Earth (even more
depending on the hyperbolic escape velocity vector).  
In Figure  \ref{fig:angles} we show the angles at which these forces
would act for a hyperbolic trajectory in the ecliptic, between 20 and 40 AU.
The eccentricity is 5 and the craft travels at approximately a terminal 
velocity of 5 AU/yr.  ($a$, the minimum distance from the hyperbola to
its intersecting asymptotes, is 1.56 AU.)  The reference 
curve (1) at zero degrees is the constant direction
towards the Sun.  Other angles are in reference to this. 
Starting to the right in the plane for definiteness, the angle
towards the Earth (2) is a cosine curve which is modified by an $1/r$
envelope as the craft moves further out.  The angle from the Sun
to the trajectory line 
is shown in (3).  Finally, the direction along the spin axis (4) is a
series of decreasing step functions.  This indicates two maneuvers per
year to place the antenna direction between the maximum Earth
direction and the null Sun direction, performed as the Earth passes
from one side of the Sun to the other.  

Looking more closely, it turns out 
that navigation alone can give evidence to help distinguish among
the directions of interest. 
 
A pair of micro-radian quality pointing sensors  (for both pointing
control and also stability -- now standard in the field) will enable one
to position the spacecraft with respect to inertial standards of rest
to a very high accuracy. The use of 3-D navigation discussed above
will result in a precise spacecraft positioning with respect to the
solar system barycentric reference frame. As with the Pioneers, the
accuracy of the determination will depend on the properties of the
antenna radiation pattern. Highly pointed radiation patterns are
available for  higher communication frequencies. In order to
be on the safe side, one can use a standard X-band antenna with a
0.5$^\circ$ angular resolution.  Therefore, 
if the anomaly is directed towards the Sun (1), 
a combination of the above two methods will be able 
to establish such a direction with sufficiently high accuracy. 
 
If the anomaly is directed towards the Earth (2), the current accuracy of
the Earth's ephemerides will be a key to determining this fact.
Furthermore, in this case one would clearly see a dumped
sinusoidal signal that is characteristic to this situation (see above and
Figure \ref{fig:angles}). The use of standard hardware discussed
above will enable one to accurately establish this direction with a high
signal-to-noise ratio.  

Further, an almost-linear angular change approaching the direction towards the
Sun (also highly correlated with the hyperbolical trajectory) would
indicate a trajectory-related source for the anomaly (3). This situation
will be even more pronounced if the spacecraft were to perform a
planetary fly-by. In the case of a fly-by, a sudden change in the
anomaly's direction will strongly suggest a trajectory-related
source for the anomaly. 

Finally, a step-function-like behavior of the
anomaly, strongly correlated with the maneuver history, would 
clearly support
any anomaly directed along the spin-axis (4).  
As a result, a combination of the standard navigation methods
addressed above in combination of the symmetric spacecraft design
(discussed in Sec. \ref{heatsymmetry})  would enable one to
discriminate between these four different directions of the anomaly
with a sufficiently high accuracy.  

It is clear that these four possible anomaly directions all have
entirely different characters.  The proposed mission is being designed
with this issue in mind. 
The use of antennas with highly pointed radiation patterns and of star 
pointing sensors, creates even better
conditions for resolving the true direction of the anomaly then does   
the use of standard navigation techniques alone. 
On a spacecraft with these additional capabilities, 
all on-board systematics will become a common mode 
factor contributing to all the attitude sensors and antennas. 
The combination of all the attitude measurements will enable 
one to clearly separate the effects of the on-board systematics 
referenced to the direction towards the Sun (1). 

This relaxes the requirement on the accuracy of the 3-D spacecraft 
navigation, allowing it to be as large as 
$0.01 \times 10^{-8}$ cm/s$^2$, 
as seen from the solar system barycentric reference frame.
In Sec.~\ref{sec:methodology} we  determine that the
expected 3-D navigational accuracy will be on the order of
$0.003 \times 10^{-8}$ cm/s$^2$.
At this resolution, the main features of the signatures of 
Fig. \ref{fig:angles} can be distinguished over a year.

This is one of the ways our mission navigation will also provide evidence 
on the origin of the anomaly, by helping to determine its direction. 
It will help to decide between the few possible alternative
mechanisms and physical causes for the anomaly. 
This answer will be important
in the more general frameworks of the solar system
ephemerides as well as spacecraft design and navigation.  

In Section \ref{antennas} we described the rotation test that will
definitively differentiate between an internal
systematic origin for the anomaly and all external origins.  
For this differentiation the 
navigational test is a backup to the rotation test.  But it will
provide information on the spatial direction that the rotation test
does not.


\section{Expected Accuracy}
\label{accuracy} 

The current mission plan calls for a nominal mission life time of 7
years (see Table \ref{tab:summary}). In the initial 3 years  the
spacecraft will reach a distance of at least 15 AU, where the data
will begin to become clear of the solar radiation bias and hence
will be of most importance to our investigation. We have the remaining
4 years of the nominal mission life time to conduct the investigation
using this cleaner data. 
 

\subsection{Data Quality}
\label{sec:methodology}

Both ranging and Doppler data will be used to achieve the required
sensitivity level for small accelerations.  

As with the Pioneers, the 
Doppler data could also be time differentiated in batches over days or
months in order to obtain independent averages of acceleration at a
sample interval equal to the batch interval. With this approach, the
standard error, $\sigma_a$, for the reduced acceleration data set is
proportional to the Allan variance \cite{old3}, $\sigma_y$, for the
fractional Doppler frequency ($y = \Delta \nu/\nu$) at 1000 seconds 
integration time. The proportionality constant is roughly $c/\tau$,
where $\tau$ is the sample interval for the acceleration data. 
The relation $\sigma_a=c\sigma_y/\tau$ is commonly used to estimate
the expected sensitivity to small accelerations.  

Currently, most of the coherent DSN tracking for NASA missions is done
by using a standard tracking configuration with X-band ($\sim$8.4 GHz)
transmitted and transponded. It is known that by far the dominant error
source is spectral broadening of the radio carrier
frequency by the interplanetary plasma, with a corresponding increase in
Doppler noise. Because of the $1/\nu^2$ dispersive nature of the
interplanetary plasma noise, our choice of X-band results in
a factor of 10 improvement over S-band ($\sim$2.3
GHz).  Ka-band radio-tracking would produce an even better 
sensitivity to small forces as opposed to the above X-band
capability.  Ka-band tracking configuration is on its way to 
being a standard option for NASA missions in the future \cite{casradio}.  

For estimation purposes, assume an Allan variance of $\sigma_y =
3.2\times10^{-16}$ in 1000 seconds of integration time, which is an
appropriate choice for system with a combination of coherent X- and 
Ka-bands \cite{new-flyby}. This results in an expected 
acceleration sensitivity of $\sigma_a=c\sigma_y/\tau\sqrt{N}\simeq
0.019\times10^{-8}$ cm/s$^2$ for a one month sample interval, where 
$N \simeq 2.6\times10^3$ is the number of independent single
measurements of the clock with duration 1000 seconds that are
performed in one month. This means a year sample 
would yield accuracy of $\sigma_a\simeq 0.005\times10^{-8}$ cm/s$^2$.  

Furthermore, with 9 months of coherent DSN tracking each year and 4
years of potential data collection, an X-/Ka-band tracking
configuration would enable an acceleration sensitivity of 
$\sigma_a\simeq 0.003\times10^{-8}$ cm/s$^2$,\footnote{A similar
estimate using X-band alone would yield a number around 
$\sigma_a\simeq 0.01\times10^{-8}$ cm/s$^2$.}
thus increasing our data
resolution to any small forces affecting the spacecraft motion.\footnote{ 
A similar estimate can be obtained directly from the
results of the Pioneer analysis \cite{old3}, where a statistical
{\small WLS} error of  $0.01$ to $0.02 \times 10^{-8}$ cm/s$^2$ was obtained
for runs of order 3 years. In fact. one-year runs for the Pioneer
S-band data set produced a similar statistical error of $\sim 0.02
\times 10^{-8}$ cm/s$^2$.  This supports the above 
estimate obtained by using the higher frequency X-/Ka-bands.} 

Therefore, our current analysis and mission simulations indicate that the
expected 3-D navigational accuracy may be characterized by the
following sensitivities: as seen from the solar system barycentric frame: 
$\sigma_{a_P}^r= 0.003 \times10^{-8}~{\rm cm/s}^2$ 
in the ``line-of-sight" direction.   Even with  
${\Delta}$DOR the two remaining orthogonal components will be larger,  
$\sigma_{a_P}^x=\sigma_{a_P}^y \approx  0.006 \times10^{-8}~{\rm
cm/s}^2$.  However,  since 
\be
\sigma_{a_P} =\left[\left(\frac{{a_P}^r}{a_P}\right)^2(\sigma_{a_P}^r)^2
                  + \left(\frac{{a_P}^x}{a_P}\right)^2(\sigma_{a_P}^x)^2
                  + \left(\frac{{a_P}^y}{a_P}\right)^2(\sigma_{a_P}^y)^2
\right]^{1/2}
\ee
and we know that $({a_P}^x/{a_P})^2$  and $({a_P}^x/{a_P})^2$ 
are both very small compared to unity, $\sigma_{a_P}^r$ dominates
the {\small RMS} error.  

Therefore,  we have
$\sigma_{a_P} \approx 0.003 \times10^{-8}~{\rm cm/s}^2$.
Although, this is only a preliminary estimate for the potential
precision of our acceleration solution, it yields confidence in the
experimental concept. 


\subsection{Systematics} 
\label{sec:systematics}

This mission is designed not only to verify the existence of the
anomaly but also to clearly determine if the anomaly is due to
an external cause or to systematics.  
In the case of Pioneers, the on-board generated systematics were the 
largest contributors to the total error budget given in Table II of
Ref. \cite{old3}. Among the most important constituents, the radio
beam reaction force produced the largest bias to our result,
$1.10\times 10^{-8}$ cm/s$^2$. Being of opposite sign to the
measurement, it resulted in a larger
final Pioneer anomaly.  The largest bias/uncertainty was from RTG heat
reflecting off the spacecraft, 
$(-0.55 \pm 0.55) \times 10^{-8}$ cm/s$^2$.  Large uncertainties  also
come from differential emissivity of the RTGs, radiative cooling, and
gas leaks, $\pm 0.85$, $\pm 0.48$, and $\pm 0.56$, respectively,
$\times 10^{-8}$ cm/s$^2$. The least significant factors of the error
budget were those external to the spacecraft and the computational
data-handling systematics.   

With the exception of the admittedly novel second antenna and the side
louvers, much of the craft architecture is specifically taken from the
Pioneers.  This allows us to once again use the lessons from the
Pioneers in analyzing our error budget.  

"As with the Pioneers,  we have decided to treat all the errors (both
experimental and systematic)
in a least squares {\it uncorrelated} manner.\footnote{
Combining experimental and systematic errors is a  problem that is
quite common in experimental physics.   
The usual solutions are to either treat them as we have or
else to list them as two errors in sum for the result.  There are
un-rigorous arguments for both methods.}  
The constituents of the error budget are listed
separately in three different categories: 1) systematics generated
external  to the spacecraft and  2) on-board generated systematics.
The error itself is conservatively found to be a factor of 
two orders of magnitude times
smaller than the Pioneer anomaly signal. 

Finally, when compared to the
biases for the Pioneers, these biases are almost nonexistent (less
than the error).  This comes directly from the symmetry of our design
and mission concept.

The results of our analysis of the systematics of proposed design, 
given in the reminder of this subsection, are summarized and included  in
Section \ref{budget}, which serves as a large-constituent ``error
budget.'' This budget is useful both for evaluating the expected
accuracy of our solution for $a_P$ and also for guiding  possible
future efforts with other spacecraft. 


\subsubsection{External Systematics}
\label{externkal}

As we demonstrated in our previous work \cite{old3}, the external
systematics are all well characterized and mainly very small,
depending somewhat on how far out from the Sun the measurements are
done.  These small systematics  are the solar wind, solar corona
(especially with the X-/Ka-bands of this mission), electromagnetic Lorentz
forces, influence of the Earth's orientation, mechanical and phase
stability of the DSN antennae, phase stability and clocks, and the
troposphere and ionosphere \cite{old3}.  Simply because of the similar
size and mass of the craft, these effects which were small for the
Pioneers would remain small here. We estimate their combined {\small
RMS} influence would be $ \leq 0.01 \times 10^{-8}$  cm/s$^2$.  

The expected quality of the data eliminates the need for a 
more detailed analysis of the computation systematics. Further, 
the data and the models used to analyze it will introduce 
an error less than the above. (See Section \ref{sec:methodology}.)

The solar radiation pressure is a bias factor that must be accounted for.
Standardly the parameterization of the pressure effects for different
spacecraft orientations with respect to the Sun is done during
early orbit and for the Pioneers was good to better than 5\%
\cite{5percent}.  Even so, close in to the Sun 
there can be confusion between this and the
computed vs. measured constant RTG systematic, as happened with 
Cassini.\footnote{In Earth-Venus cruise, the  
Cassini orbit determination originally found a significantly different
systematic bias from the RTGs mounted on the front than had been
predicted by thermal models.  
This was later determined to have been due the problem of
disentangling the RTG systematic from the solar radiation pressure
while so close in to the Sun.  There the pressure was so much larger 
and the craft displayed varying aspects to the Sun in the flyby
 trajectory
\cite{cassiniheat2}.} 

The size of the actual solar radiation bias varies, of course,  
as the inverse of the distance from the Sun.  At 20
AU the signal  would be around  
$4 \times 10^{-8}$  cm/s$^2$ for the proposed craft.  Further, at 
that distance the craft's attitude towards the Sun would be less than
3 degrees.  This angular variation would provide 
the only component that would vary from an inverse square fall off 
and so it becomes vanishingly small.  
Although there might be some uncertainty (a few kg) in the total mass
propellant consumed, at this time very little more would be used.  The
variation over a year would be less that 1 kg ($\sim 0.3$~\% of the craft's
mass), or a signal varying from $1/r^2$ by a factor of 0.003.
But both of these modifications are still on top of the $1/r^2$ 
variation which at 20 AU  decreases by almost 10\% in 1 AU.  If the
craft were going 5 AU/yr, the signal would decrease 36\% in a year. 
Since at this level the signal strength can be determined to about a
part in a thousand, we can use our great distance from the Sun to 
place a bound on the signal's {\small RMS}  error of 
${0.02 \times 10^{-8}}$  cm/s$^2$.  

The only other significant systematic in this category would be from
some uncalculated gravity effect, the most likely being from  
the Kuiper belt.  Since the gravitational  force is inertial, the same
bound can be used as that for the Pioneer craft \cite{old3}.  
There it was shown that this possibility is limited to 
$0.03 \times 10^{-8}$ cm/s$^2$.  
Although the galactic field is of the size of the Pioneer anomaly, 
it too  can not be the origin
of the anomaly. This is because of the fact that Pioneer 11 was
traveling roughly in the direction of the solar system's motion within
the galaxy and Pioneer 10 was moving almost in the opposite
direction. Further, a galactic tidal force also can not explain 
the anomaly \cite{old3}. 

As noted in the introduction, 
the Pioneer anomaly could have been caused by an interplanetary
density of $4 \times 10^{-19}$ g/cm$^3$, that would have to be
constant over 50 AU (something hard to understand). Therefore, since this
craft will be going at a higher velocity than the Pioneers were and
have roughly the cross-sectional area and mass of the Pioneers, the
resultant effect would be a factor of $\sim v^2$ higher
\cite{saildrag}.  Further, 3-dimensional tracking would help show 
if the acceleration was in the direction of motion of the craft
(see Section \ref{navigation}), giving an indication 
if this were the origin of the anomaly.


\subsubsection{On-board Generated Systematics}
\label{on-board}

It is here that our design makes the most significant contribution to
error reduction.  On-board generated systematic contributed the most
significant parts of the total Pioneer error budget of $(+0.90 \pm
1.33) \times 10^{-8}$  cm/s$^2$. Because of the rotational cylindrical
symmetry of the craft, on-board systematics like the radio-beam
reaction force, RTG 
heat reflection, differential emissivity of the RTGs non-isotropic
radiative cooling of the craft, or expelled helium from the RTGs  
could only contribute fore or aft along the spin-axis.  

As we now argue, the additional fore/aft symmetry of the current
design should limit the effect of any remaining asymmetry to 
${0.03 \times 10^{-8}}$  cm/s$^2$.  This is first because of the small
size of any systematics to begin with.  However, even more importantly,
they can be canceled down to near the navigation data error 
by the $180^\circ$ rotation maneuver described in Section \ref{antennas}. 

For example, what if there is an imperfection in the planned
symmetry, such as some of the louvers stick, one of the radio beams
does not emit properly, or the two antennas are not equivalent?  Well
then, after obtaining a ``forward'' measurement  for the
anomaly of $a_{f1}$,
when one turns the craft around and measures a new
``backwards'' $a_{b1}$, one knows that the anomaly, $a_P$, and the
bias caused by the asymmetry, $\Delta$, are  
\be
a_P = \lfrac{1}{2}(a_{f1}+ a_{b1}), ~~~~~~ 
            \Delta =  \lfrac{1}{2}(a_{f1}- a_{b1}).
\ee
This measurement is limited only by the inaccuracy of the mass
determination from fuel usage of about 1/3\% craft mass in a year  
(and the radio frequency measurement error).
Normalized to the Pioneer results, multiplying this factor yields   
an error of ${0.03 \times 10^{-8}}$ cm/s$^2$, which we take to our   
fore/aft internal asymmetry error. 

The gas leakage error is the hardest to deal with.   Small gas leaks that
are usually negligible for other missions could in principle cause a
problem here.  

For the Pioneers
there were anomalous spin-rate changes that could be correlated with
changes of the exact values of the short term $a_P$.  
The correlations between the spin-rate changes and $a_P$ 
were good to
$0.2 \times 10^{-8}$  cm/s$^2$ and better.\footnote{But even so, a
conservative error of $0.56 \times 10^{-8}$  cm/s$^2$ was 
quoted \cite{old3}.} 

Here we will use modern monitoring to attentively follow the
spin-rate changes so as to improve on the observed Pioneer 
correlations.  Further, we will be aided  by the current 
technology development of thrusters (especially the development of
$\mu N$-thrusters for the LISA mission).\footnote{Consideration can also
be given to double-valves that have an escape that is directed along 
the rotation of the craft.}  A final capability is that we will rotate 
the space craft a first, second, and even more times (say one year
runs).  The initial value (facing forward) will be $a_{f1}$.  After
the first rotation a (backward facing) value, $a_{b1}$, will be
obtained; and similarly thereafter $a_{f2}$, $a_{b2}$, etc. 
Any differences among the $a_{fi}$'s and  $a_{b2}$'s will be a
measure of the error introduced by gas leaks.  This all leads to an 
error of ${0.04 \times 10^{-8}}$  cm/s$^2$.


\subsection{Mission Error Budget}
   \label{budget}

The results of the previous discussions can be seen in  
in Table \ref{errorbudget}, which gives a summary the 
significant contributions o the error budget. 
Adding these errors as 
{\small{RMS}} one obtains a final error limited to  
$0.06 \times 10^{-8}$ cm/s$^2$.  
{\it We emphasize that this is with no bias!}  
It is a factor of more than a 100 smaller than the Pioneer anomaly! 
Even with existing technologies, his mission would leave no doubt as
to the existence of the anomaly.  

This error is how well one will be able to determine the
size of the signal, the ``accuracy'' determined by 
all the (mainly systematic) errors.  This is different than the 
``precision'' of the measurement, which is determined by the statistics.
The precision will be smaller, as we discussed in Section 
\ref{sec:methodology}. 


\begin{table*}[ht!]
\begin{center}
\caption{Summary of {\it Significant} Error Budget Constituents.
\label{errorbudget}} \vskip 20pt
\begin{tabular}{rlll} \hline\hline
Item   & Description of error budget constituents & 
 Uncertainty &    $10^{-8} ~\rm cm/s^2$  \\\hline
   &                 &             &         \\
 1 &  {\sf Three-dimensional acceleration uncertainty from data}  
            &&  $\pm 0.003$ \\
  [10pt]
 2 & {\sf Systematics generated external  to the spacecraft:}  && \\
   & a) Solar radiation pressure          &     & $\pm 0.02$\\
   & b) Influence of the Kuiper belt's gravity     && $\pm 0.03$ \\   
 [10pt] 
 3 & {\sf On-board generated systematics:} &&         \\
   & a) Fore-aft asymmetry (heat, radio)   && $\pm 0.03$ \\
   & b) Gas leakage           &         &  $\pm 0.04$   \\
[10pt] 
\hline
   &                              &&         \\
   & Estimate of total error     && $\pm 0.06$       \\
   &                              &&         \\
\hline\hline
\end{tabular} 
\end{center} 
\end{table*}


\baselineskip=.165in

For context, return to the Pioneer analysis \cite{old3}.  
The ``experimental'' number obtained is 
$a_{P({\tt exper)}}= (7.84 \pm 0.01)~\times~10^{-8}~\mathrm{cm/s}^2$.   
This statistical error in the data, $0.01~
\times~10^{-8}~\mathrm{cm/s}^2$ is very small. That the anomaly 
is in the data, was independently verified \cite{craig}.  The precision
of the data, there and also here, is not the question. 

The Pioneer analysis also 
found \cite{old3} that the systematic {\it bias and
error} are.  $(+0.90~\pm~1.33)\times~10^{-8}~\mathrm{cm/s}^2$. 
This then led to the final number, 
$a_P = (8.74 \pm 1.33) \times 10^{-8} ~~{\rm cm/s}^2$.  
Seemingly, this more than 6$\sigma$ result would appear to be
sufficient.  

But none-the-less, the main question has become whether or not 
in some manner or other, systematics could have caused the
anomaly anyway.  That is, although the calculated bias was 
$+0.90~\times~10^{-8}~\mathrm{cm/s}^2$, nonetheless,
in some not understood way, could the anomaly have been due to
systematics?

The mission proposed here, because of the symmetry of the craft,
attitude control, navigational capability and the mission design, will
have a low {\it conservative} systematic error of 
$ 0.06 \times 10^{-8} ~~{\rm cm/s}^2$ or 
less, which error cannot be assailed.
Therefore, this mission will absolutely settle the above question.



\section{A Summary and the Future} 
\label{sec:concl}

In Table \ref{tab:summary} we summarize the properties of our mission.   
This table puts together the various pieces of the program
that we have described.  With this program the origin of the Pioneer
anomaly can be unambiguously determined. 
In addition, independent of whether a Pioneer Anomaly test will be a
stand alone experiment or a probe from a larger mission, we have
shown here that such a mission is feasible and also defined its
requirements.  Any mission concept that may ultimately be chosen will need 
to be guided by these principles.


\def\ref{\vskip 0pt \par \hangindent 12pt\noindent}
\vskip 2pt 
\begin{table*}[h!]
\caption{Mission Summary \label{tab:summary}}
\vskip 5pt
\begin{tabular}{|c|m{13.0cm}|} \hline 
&\\[-12pt]
Objectives & 
\ref$\bullet$\hskip 6pt
To search for any {\it unmodeled} small acceleration affecting the spacecraft 
motion at the level of ${\sim 0.1 \times 10^{-8}}$ cm/s$^2$ or less.  
\ref$\bullet$\hskip 6pt 
Determine the physical origin of any anomaly, if found. 
\\[3pt]\hline
&\\[-12pt]
Features  & 
\ref$\bullet$\hskip 6pt
A standard spacecraft bus that allows thermal louvers to be on the
sides to provide symmetric fore/aft thermal rejection.  
\ref$\bullet$\hskip 6pt 
Fore/aft symmetric  design with twin antennae (``yo-yo'' concept).
\\[3pt]\hline
&\\[-12pt] 
Orbit & 
\ref$\bullet$\hskip 6pt
Solar system escape trajectory -- possibly in the plane of ecliptic,
co-moving with the solar system's direction within the galaxy. 
\ref$\bullet$\hskip 6pt 
Spacecraft moving with a velocity of 5 AU or more per year, reaching
15 AU in 3 years time or less. 
\\[3pt]\hline
&\\[-12pt]
Launcher  &
\ref$\bullet$\hskip 6pt
Delta IV 2425 or any heavy vehicle, i.e., Proton, Ariane V.
\\[3pt]\hline
&\\[-12pt]
Spacecraft  &  
\ref$\bullet$\hskip 6pt
Power at launch: $\sim$200W provided by RTGs located on  booms at a
distance of $\sim$3 m from the rotational axis of the spacecraft.  
\ref$\bullet$\hskip 6pt
Redundancy: single-string.
\ref$\bullet$\hskip 6pt
Mass: s/c dry $\sim$300 kg; propellant $\sim$40 kg;  
total  launch  $\sim$500 kg. 
\ref$\bullet$\hskip 6pt
Dimensions at launch: diameter $\sim$2.5 m; height: $\sim$3.5 m {\it including
both Cassegrain antennae}. 
\ref$\bullet$\hskip 6pt
Attitude control: spin-stabilized spacecraft. 
\ref$\bullet$\hskip 6pt
Navigation: Doppler,
range,  and possibly VLBI and/or  $\Delta$DOR. 
\ref$\bullet$\hskip 6pt
Pointing: control  6 $\mu$rad; knowledge 3 $\mu$rad;
stability 0.1 $\mu$rad/sec. 
\ref$\bullet$\hskip 6pt
Telemetry:  rate  1 Kbps.
\\[3pt]\hline
&\\[-12pt]
 Lifetime&  
7 years (nominal for velocity of 5 AU/year); 12 years (extended).\\
 & 5 years (nominal for velocity of 10 AU/year); 8 years (extended).
\\[3pt]\hline  
\end{tabular}
\end{table*}


\baselineskip=.165in

The determination of the anomaly will be of great 
scientific interest and value.
Even if, in the end, the anomaly is due to some
systematic, this knowledge will greatly aid future mission design
and navigational programs.  But if the anomaly is due to some
not-understood physics, the importance would be spell-binding.  The
benefits to the community from this program would then be 
enormous.

But addressing this specific problem has motivated thought on a more
general one.  Our knowledge of the dynamical metrology of the outer
solar system is relatively very poor.  The Pioneers yielded the best
measurements we have for deep-space hyperbolic orbits.  
These measurements were imprecise compared to what we could
hope for today.  

Our mission will use already existing navigation  
methods to conduct the most precise spacecraft navigation ever 
performed in deep space.  It will rely on a novel spacecraft 
design to minimize the effects of small forces acting on the craft from 
both external and internal causes.  It will have a navigational
accuracy two orders of magnitude better than that currently
available.  With its advanced spacecraft design and operations, 
the mission will reduce sysematics to an unprecidented level.  
This will help develop the critical expertise that wil be needed to 
create the low-noise environments needed for precision 
space deployments of the 21st century.  Thereby it will allow special
tests of fundamental physics in space. All required technologies have
already been demonstrated. This mission has no analogs: it is a unique
natural extension of precise gravitational experiments in the solar
system.


\section*{Acknowledgments}

M.M.N. acknowledges support by the U.S. DOE.
The work of S.G.T was performed at the
Jet Propulsion Laboratory, California Institute of Technology, under
contract with the  National Aeronautics and Space Administration.  



\end{document}